\documentstyle[12pt,epsf]{article}
\newcommand {\beq}{\begin{equation}}
\newcommand {\eeq}{\end{equation}}
\newcommand {\beqa}{\begin{eqnarray}}
\newcommand {\eeqa}{\end{eqnarray}}
\newcommand {\beqan}{\begin{eqnarray*}}
\newcommand {\eeqan}{\end{eqnarray*}}
\newcommand {\n}{\nonumber \\}

\newcommand {\eq}[1]{eq.~(\ref{#1})}

\newcommand {\del}{\partial}

\begin{document}

\begin{titlepage}
\renewcommand{\thefootnote}{\fnsymbol{footnote}}

\begin{flushright}
\begin{tabular}{l}
KEK-TH-644 \\
NSF-ITP-99-105\\
\end{tabular}
\end{flushright}

\vspace{2cm}
\begin{center}
         {\Large Branched Polymer Revisited} \\
\end{center}
\vspace{1cm}

\begin{center}

           Hajime A{\sc oki}$^{1)}$\footnote
           {
e-mail address : haoki@itp.ucsb.edu, JSPS abroad research fellow},
           Satoshi I{\sc so}$^{2)}$\footnote
           {
e-mail address : satoshi.iso@kek.jp},
           Hikaru K{\sc awai}$^{3)}$\footnote
           {
e-mail address : hkawai@gauge.scphys.kyoto-u.ac.jp}
          and Yoshihisa K{\sc itazawa}$^{2)}$\footnote
           {
e-mail address : kitazawa@post.kek.jp}\\
\vspace{1cm}
        $^{1)}$ {\it Institute for Theoretical Physics, UCSB}\\
               {\it Santa Barbara, , CA 93106, USA} \\
        $^{2)}$ {\it High Energy Accelerator Research Organization (KEK),}\\
               {\it Tsukuba, Ibaraki 305, Japan} \\
        $^{3)}$ {\it Department of Physics, Kyoto University,} \\
                 {\it Kyoto 606-8502, Japan}\\
\end{center}

\vfill

\begin{abstract}
\noindent
We show that correlation functions for branched polymers correspond
to those for $\phi^3$ theory with a single mass insertion,
not those for the $\phi^3$ theory themselves, as has been widely believed.
In particular, the two-point function behaves as $1/p^4$, not as $1/p^2$.
This behavior is consistent with the fact that the Hausdorff dimension
of the branched polymer is four.
\end{abstract}
\vfill
\end{titlepage}
\vfil\eject

\section{Introduction}
\setcounter{equation}{0}
\setcounter{footnote}{0}
Branched polymers are the simplest generalization of the
random walk and have been studied extensively.
It is of great importance not only in statistical physics but
also in particle physics, in particular, for understanding
the critical behavior of random surface models and quantum gravity
(see, for example, \cite{ambj}).
In the paper \cite{AIKKT}, we have studied the dynamics of the type
IIB  matrix model of a constructive formulation of superstring
(\cite{IKKT} and see \cite{review} for review).
In our matrix model approach, the eigenvalues of matrices
are interpreted as space-time coordinates.
In these investigations, we find the system of branched polymers
in a simple approximation.
Although it is far from the flat four dimensional manifold, branched
polymers share the same (fractal) dimensionality four with our space-time.
It might be the first indication that superstring can explain the
dimensionality
of our space-time.
\par
In this  letter, we comment on a field theoretic description
of branched polymers.
It is well-known
that a system of random walks is described by a free scalar field
theory if there is no  effect of self-avoidance.
Similarly it is widely believed that the system of branched polymers
becomes a scalar field theory with a three-point coupling,
that is, $\phi^3$ scalar field theory
(see, for example, \cite{ambj}).
We will, however,  show that it is not so by treating the
universal part of the partition function carefully.
The system of branched polymers without self-avoidance can be
exactly solvable by introducing the grand canonical ensemble
and using the so called Schwinger-Dyson technique.
In order to extract the correct large $N$ limit ($N$ is the system size)
or the thermodynamic limit, we have to check
that the grand canonical ensemble is dominated
by the larger size system.
In other words, we have to extract the universal part.

On the contrary to a  naive summation which
makes us conclude that the system of branched polymers
is described by a $\phi^3$ scalar field theory,
a careful treatment shows that we need a single
mass insertion in each n-point function
of the $\phi^3$ scalar field theory.
Mass insertion here means a change of a propagator
in each n-point function from
an ordinary one, $1/p^2$, to $1/p^4$.
In particular, the two-point function behaves as $1/p^4$, not $1/p^2$.
Let us count the number of points which lie within distance $R$ from a fixed
point
in $D(>4)$ dimensions.
In the random walk, it can be estimated as $R^2$ by using the two point
function.
We obtain $R^4$ for branched polymers by using the $1/p^4$ type propagator.
So our finding are consistent with the claim that branched polymers are four
dimensional fractals.
A multi-point Green function is given by a sum of graphs of the
corresponding Green function
for $\phi^3$ scalar field theory at tree level with
a single mass insertion in each graph.
\par
The organization of this paper is the following.
First we define a canonical ensemble for a system of branched polymers
(sec. 2.1) and then introduce grand canonical ensembles (sec. 2.2).
We emphasize that a definition of grand canonical ensemble is not unique.
In sec. 2.3, we solve Schwinger-Dyson equations for the
grand canonical ensembles  and obtain  naive results
of the correlation functions of a scalar $\phi^3$ theory.
In sec. 2.4, we consider the thermodynamic limit
and give the correct universal behavior of the correlation functions.
 Finally in section 3, we conclude and give an interpretation
of our result.

\section{Branched polymer dynamics}
\setcounter{equation}{0}
\subsection{Canonical Ensemble}

\begin{figure}[b]
\begin{center}
\leavevmode
\epsfxsize=6cm
\epsfbox{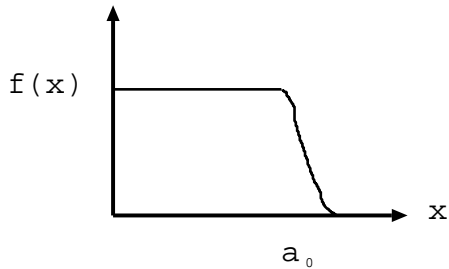}
\caption{Form of the function $f(x)$.
It damps sufficiently fast at long distance compared to the typical
length scale $a_{0}$.
}
\label{fig:fx}
\end{center}
\end{figure}

Branched polymers are a statistical system of
$N$ points connected by $N-1$ bonds whose lengths are of order
$a_{0}$.
The canonical  partition function
is defined as
\beq
Z_N=\frac{1}{N!}\sum_{G:{\rm tree\;\; graph}} \int \prod_{i=1}^{N} dx^i
\prod_{(ij):{\rm bonf\;\;of\;\;}G} f(x^i-x^j),
\label{eq:cpf}
\eeq
where $f(x)$ is a function assigned to each bond in each graph, and
it damps sufficiently fast at long distance compared to the typical
length scale $a_{0}$.
(See, for example, figure \ref{fig:fx}.)
The presence of the factor $1/N!$ is  due to the fact that the $N$ points
are regarded
identical.

We can calculate a partition function for each $N$,
by counting the number of all possible tree graphs:
\beqa
Z_1&=& V \\
Z_2&=& \frac{1}{2!}\hat{f}(0)V \\
Z_3&=& \frac{1}{3!}3\hat{f}(0)^2 V \\
Z_4&=& \frac{1}{4!}16\hat{f}(0)^3 V \\
   &\vdots&\n
Z_N&=& \frac{1}{N!}N^{N-2}\hat{f}(0)^{N-1} V ,
\label{eq:Zn}
\eeqa
where $\hat{f}(p)$ is a Fourier transform of $f(x)$;
\beq
\hat{f}(p) = \int dx\;\; e^{ipx}\;\; f(x),
\eeq
and $V$ is the total volume of the system.
A derivation of the  general form (\ref{eq:Zn}) is given
in the appendix.
\par
We define an (unnormalized) 
m-point correlation function of density operators as
\beqa
&&G^m_N(x^1,\cdots,x^m) \n
&=&<\rho(x^1)\cdots\rho(x^m)>_N \n
&=&\frac{1}{N!}\sum_{G:{\rm tree\;\;graph}} \int \prod_{i=1}^{N} dx^i
\prod_{(ij):{\rm bond\;\;of\;\;}G} f(x^i-x^j)\;\;\rho(x^1)\cdots\rho(x^m),
\label{eq:ccf}
\eeqa
where the density operator is defined by
\beq
\rho(x) = \sum_{i=1}^N \delta^{(d)}(x-x^i).
\eeq

\subsection{Grand canonical ensemble}
We then define  partition functions and  m-point correlation functions
in the grand canonical ensemble in the following equations:
\beqa
Z_{\kappa_0,l}&=& \sum_{N=1}^{\infty}N^l \; \kappa_0^N \; Z_N,
\label{eq:gcpf}\\
G^m_{\kappa_0,l}(x^1,\cdots,x^m)
&=& \sum_{N=1}^{\infty}N^l \; \kappa_0^N \; G^m_N(x^1,\cdots,x^m),
\label{eq:gccf}
\eeqa
where $\kappa_0$ is  fugacity.
There exist
various ways of defining  grand canonical ensemble
corresponding to the freedom to choose different weights for each fixed $N$
sector.
Here, we assigned an extra $N$-dependent factor, $N^l$,
 in addition to the usual one, $\kappa_0^N$.
The criterion for a "good" grand canonical ensemble is such that
we can take the correct thermodynamic limit, or, in other words,
we can correctly take the  universal part in the sum
(\ref{eq:gcpf}) or (\ref{eq:gccf}).
That is, the correlation functions in the grand canonical ensemble at the
critical value of fugacity should reproduce
those in the canonical ensemble for large N:
\beq
\lim_{N \to \infty } \frac{G^m_N}{Z_N}
= \lim_{\kappa_0 \to \kappa_{0,c}}
\frac{G^m_{\kappa_0,l}}{Z_{\kappa_0,l}}.
\eeq

We illustrate the above mentioned criterion by taking partition function
as an example.
Since the canonical ensemble partition function (\ref{eq:Zn})
behaves at large N as
\beq
Z_N \sim \frac{N^{-5/2}}{\sqrt{2\pi} e^{-N}} \hat{f}(0)^{N-1} V,
\eeq
the grand canonical partition function is approximated by
\beqa
Z_{\kappa_0,l}&\sim& \frac{V}{\sqrt{2\pi} \hat{f}(0)}
       \sum_{N=1}^{\infty} N^{l-5/2} \;\; (\kappa/\kappa_c)^N \\
&\sim& \frac{V}{\sqrt{2\pi} \hat{f}(0)}
\int_0^{\infty} dN \;\; N^{l-5/2} \;\; e^{-N \Delta \kappa/\kappa_c},
\label{eq:integrand}
\eeqa
where
\beqa
\kappa&=&\hat{f}(0)\kappa_0, \\
\kappa_c &=& e^{-1},
\label{eq:kc1}\\
\Delta\kappa&=&\kappa_c - \kappa.
\eeqa

If we take $l$ sufficiently large,
the integrand in \eq{eq:integrand} has a peak at
$N \sim \kappa_c/\Delta \kappa$ and
 we can make N large by approaching $\kappa$ to $\kappa_c$.
On the other hand, if $l$ is not sufficiently large, a non-universal
small $N$ behavior dominates the summation and we cannot obtain
the  correct answer of the large $N$ limit by a grand canonical ensemble.

\subsection{Schwinger-Dyson eq.}
In this subsection, we recapitulate the (naive) arguments that the correlation
functions for branched polymers are given by massless $\phi ^3$ theory
\cite{ambj}\cite{AIKKT}.
Let's consider the following correlation functions,
which are suitable for a
Schwinger-Dyson analysis:
\beqa
F^m(x^1,\cdots,x^m) =&&
\sum_{N=1}^{\infty} \frac{\kappa_0^N}{(N-m)!}
\sum_{G:{\rm tree\;\;graph}}
\int \prod_{i=1}^{N} dy^i \;\;
\delta^{(d)}(y^1-x^1) \cdots \delta^{(d)}(y^m-x^m)\n
&&\times \prod_{(ij):{\rm bond\;\;of\;\;}G} f(y^i-y^j).
\label{eq:sdcf}
\eeqa
The factor $\frac{1}{(N-m)!}$ means
that $(N-m)$ points other than the fixed $m$ points are regarded identical.
We can see from  \eq{eq:gccf} and \eq{eq:ccf} that
\beq
F^m(x^1,\cdots,x^m)
\simeq G^m_{\kappa_0,l=0}(x^1,\cdots,x^m)
\label{eq:fandg}
\eeq
in the large N limit.
We write a Fourier transform of $F^m(x^1,\cdots,x^m)$
as $\hat{F}^m(p^1,\cdots,p^{m-1})$:
\beqa
&&(2\pi)^d \delta^{(d)}(p^1+\cdots+p^m)
\hat{F}^m(p^1,\cdots,p^{m-1}) \n
 && =
\int d^d x^1 \cdots d^d x^m \;\;e^{i p^1 x^1} \cdots e^{i p^m x^m}
\;\;F^m(x^1,\cdots,x^m).
\eeqa

\begin{figure}[h]
\begin{center}
\leavevmode
\epsfxsize=5cm
\epsfbox{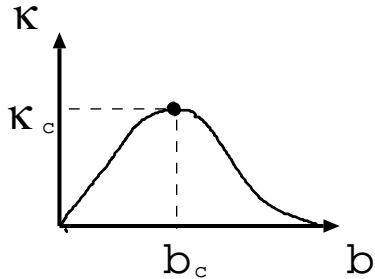}
\caption{Schwinger-Dyson equation, $\kappa = b e^{-b}$.
At the critical point, $b_c=1, \kappa_c=e^{-1}$.
}
\label{fig:bk}
\end{center}
\end{figure}

Schwinger Dyson equation for 1-point function $\hat{F}^1$ becomes
\beq
b  = \kappa e^b,
\label{eq:sdb}
\eeq
where
\beq
b \equiv \hat{f}(0) \hat{F}^1,
\label{eq:defb}
\eeq
as can be seen from figure \ref{fig:sd}.
Figure \ref{fig:bk} illustrates \eq{eq:sdb}.

\begin{figure}[h]
\begin{center}
\leavevmode
\epsfxsize=6cm
\epsfbox{sd.eps}
\caption{Schwinger-Dyson equation for one point function.
A grey blob and a black point mean $\hat{F}_1$ and
$\kappa_0$, respectively.
}
\label{fig:sd}
\end{center}
\end{figure}

At the  critical point,
\beqa
b_c &=& 1, \\
\kappa_c &=& e^{-1},
\label{eq:kc2}
\eeqa
1-point function $\del b/\del k$ diverges.
Near this critical point, $N$ becomes large;
\beq
\Delta b \simeq \sqrt{2e} \sqrt{\Delta \kappa} \sim 1/\sqrt{N},
\label{eq:delbsqrtn}
\eeq
where
\beqa
\Delta b =b_c-b, \\
\Delta \kappa =\kappa_c-\kappa.
\eeqa
\par
\begin{figure}[h]
\begin{center}
\leavevmode
\epsfxsize=7cm
\epsfbox{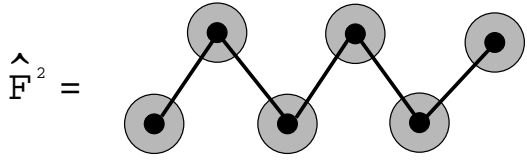}
\caption{2-point function $\hat{F}^2$ is made out of
1-point function $\hat{F}^1$, which is written by a gray blob.
}
\label{fig:f2}
\end{center}
\end{figure}

Next, we consider the 2-point function $\hat{F}^2(p)$.
When we pick up any two points on a tree graph,
we can fix the path connecting these two points.
Thus, as can be seen from figure \ref{fig:f2}, 2-point function is
calculated to be
\beqa
\hat{F}^2(p)
&=& \sum_{s=1}^{\infty} \hat{f}(p)^s(\hat{F}^1)^{s+1} \\
&=&\frac{b^2 h(p)}{\hat{f}(0)(1-b h(p))},
\eeqa
where
\beqa
h(p)&\equiv& \hat{f}(p)/\hat{f}(0)\\
    &=&1-c a_0^2 p^2+\cdots.
\eeqa

Here $c$ is a positive constant of order one.
Recall that $f(x)$ damps rapidly out of the region $0<x<a_0$,
as shown in figure \ref{fig:fx}.

Near the critical point, $b \sim b_c=1$, 2-point correlation
function behaves as
\beqa
\hat{F}^2(p) &\sim& \frac{1}{\Delta b + c a_0^2 p^2}\\
             &\sim& \frac{1}{N^{-1/2} + c a_0^2 p^2}.
\eeqa
Here we used \eq{eq:delbsqrtn}.
Thus, the correlation length is $\xi = a_0 N^{1/4}$, which implies
the Hausdorff dimension of branched polymer is four.
If the relevant length scale is shorter than the correlation length,
$\hat{F}^2(p) \sim 1/p^2$ behaves like a propagator of a massless scalar field.
Let us consider the following region:
\beq
a_0 \ll x \ll  \xi= a_0 N^{1/4}.
\label{eq:region}
\eeq
$a_0 $ gives an ultraviolet cut-off whereas
$\xi$ gives an infrared cut-off length over which correlation functions
damp rapidly.
Note that, in this region and near the critical point,
the following inequality holds:
\beq
1 \ll \frac{1}{1-bh(p)} \ll\frac{1}{1-b}.
\label{eq:ineq}
\eeq

\begin{figure}[b]
\begin{center}
\leavevmode
\epsfxsize=8cm
\epsfbox{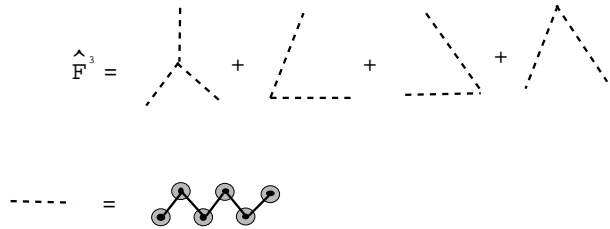}
\caption{3-point function $\hat{F}^3$ is a summation of tree diagrams
whose propagators are $\hat{F}^2$.
}
\label{fig:f3}
\end{center}
\end{figure}

Finally, we consider  correlation functions of $m>2$.
As in the case of the two-point function, $m$ points are fixed on each tree
graph.  We can uniquely fix the path connecting them on each graph.
Therefore, an $m$-point function $\hat{F}^m$ is represented as a summation over
all tree diagrams with $m$ fixed points in which
$\hat{F}^2$ appear as propagators.
For example,
\beqa
\hat{F}^3(p,q)
&=&(b/\hat{f}(0))^{-2} \hat{F}^2(p)\hat{F}^2(q)\hat{F}^2(p+q) \n
  &&+(b/\hat{f}(0))^{-1}[\hat{F}^2(p)\hat{F}^2(q)
+\hat{F}^2(p)\hat{F}^2(p+q)
+\hat{F}^2(q)\hat{F}^2(p+q)],
\label{eq:santen}
\eeqa
as we can see from figure \ref{fig:f3}.
However, because of the inequality (\ref{eq:ineq}),
the diagrams with the maximum number of propagators dominate.
In the case of 3-point function, the first term in
\eq{eq:santen} dominates.
In general, we obtain a naive result for $m$-point correlation functions;
\beq
\hat{F}^m \sim \mbox{correlation functions of massless $\phi^3$ theory at
tree level}.
\eeq
In the next subsection, we will see that these naive results
do not correspond to the correct thermodynamic results

\subsection{Correlation functions in thermodynamic limit}
As we mentioned in \eq{eq:fandg}, $\hat{G}^m_{\kappa_0,l=0}$ is
equal to $F^m$ in the  large N limit:
\beq
\hat{G}^m_{\kappa_0,l=0}(p^1,\cdots,p^{m-1})
\simeq \hat{F}^m(p^1,\cdots,p^{m-1}).
\eeq
Then, let us consider the  $m$-point correlation functions with $l \ge 1$.
>From the definition (\ref{eq:gccf}), they can be obtained
by applying $l$-th derivative to the $l=0$ case:
\beqa
\hat{G}^m_{\kappa_0,l}(p^1,\cdots,p^{m-1})
&=&(\kappa_0 \frac{\del}{\del \kappa_0})^l
\hat{G}^m_{\kappa_0,l=0}(p^1,\cdots,p^{m-1})\\
&=&(\frac{b}{1-b} \frac{\del}{\del b})^l
\hat{G}^m_{\kappa_0,l=0}(p^1,\cdots,p^{m-1})
\eeqa

2-point function with $l=1$ is given by
\beqa
\hat{G}^2_{\kappa_0,l=1}(p)
&\simeq&(\frac{b}{1-b}) \frac{\del}{\del b}
(\frac{b^2 h(p)}{1-b h(p)}) \\
&\sim& \frac{1}{(1-b)(1-bh(p))^2}\\
&\sim &  \frac{1}{(1-b)} \frac{1}{p^4}.
\eeqa
Here, we have used the inequality (\ref{eq:ineq})
since we are interested in the correlation functions
in the physical region (\ref{eq:region}) near the critical point.
This behavior is different from that of $\hat{G}^2_{\kappa_0,l=0} \sim
1/p^2$.
Similarly for $l>1$,  the behavior of 2-point function becomes
\beqa
\hat{G}^2_{\kappa_0,l \ge 1}(p)
&\sim& \frac{1}{(1-b)^{2l-1}\; (1-bh(p))^2} \\
&\sim&  \frac{1}{(1-b)^{2l-1} } \frac{1}{p^4}
\eeqa
and coincides with the $l=1$ result.
Due to the inequality (\ref{eq:ineq}),
the derivative $\frac{\del}{\del b}$ is dominated to
act on
$\frac{1}{1-b}$, not on $\frac{1}{1-bh(p)}$.
Thus, their $p$-dependences  are all the same.
This is the correct thermodynamic limit.
That is, we should consider a grand canonical correlation function
with $l \ge 1$, otherwise a non-universal small $N$ behavior
affects the summation and we cannot obtain the universal result.
Note that the behavior of $\hat{G}^2_{\kappa_0,l \ge 1}(p) \sim 1/p^4$ is
consistent with the fact
that the Hausdorff dimension of the branched polymer is four.
A naive argument expected from the figure \ref{fig:f2}
is that the effect of branching can be absorbed by renormalizing
the mass. If so, the propagator behaves as that of random walks.
We discuss in section 3 why this argument is not correct.
\par
Similarly, 3-point functions become
\beqa
\hat{G}^3_{\kappa_0,l=0}(p,q) &\sim& g(p)g(q)g(p+q), \\
\hat{G}^3_{\kappa_0,l\ge 1}(p,q) &\sim&
g(p)'g(q)g(p+q)+g(p)g(q)'g(p+q)+g(p)g(q)g(p+q)',
\eeqa
where
\beqa
g(p)&=& \frac{1}{1-bh(p)} \sim \frac{1}{p^2}, \\
g(p)' &=& \frac{1}{(1-bh(p))^2} \sim \frac{1}{p^4}.
\eeqa
Note that only one propagator in a graph is replaced by $g'(p)$,
since the derivative $\frac{\del}{\del b}$ is dominated to act
on the factor $\frac{1}{1-b}$, as in the case of 2-point functions.
Therefore, for $m$-point correlation functions,
\beqa
\hat{G}^m_{\kappa_0,l=0} &\sim& \mbox{correlation functions for
$\phi^3$ theory at tree level} \label{eq:wrong} \\
\hat{G}^m_{\kappa_0,l \ge 1} &\sim&
\mbox{correlation functions for $\phi^3$ theory at tree level
with a mass insertion}
\label{eq:correct}
\eeqa
and the correct correlation function in the thermodynamic limit
should be  given by
\eq{eq:correct}, not by \eq{eq:wrong}.

As a consistency check,
the following relation between
an ($m$+1)-point function and an $m$-point function must hold:
\beq
\hat{G}^{m+1}(p^1,\cdots ,p^{m-1} ,p^m =0)
=\hat{G}^m (p^1,\cdots ,p^{m-1}).
\label{eq:mmpo}
\eeq
It actually holds because in the L.H.S. of \eq{eq:mmpo}, the special class of
diagrams dominate in which the $m$-th end point is attached to
the propagator $g'(p^m)$.
It is due to the inequality $g(p=0) \ll g(p=0)'$.
Then, it is equal to the R.H.S. of \eq{eq:mmpo}.

\section{Conclusion and discussion}

In this letter we have shown that
the correlation functions for branched polymers are given by
those for $\phi^3$ theory at tree level {\bf with a single  mass insertion}
if we correctly take the thermodynamic limit.
It is not given by those for $\phi^3$ theory at tree level themselves,
as has been widely believed.
\par
Our result can be interpreted as follows.
Since the 1-point function behaves as
\beq
G^1_N \sim N^{-3/2} (e \hat{f}(0))^{N}
\eeq
at large N (see eq. (\ref{eq:integrand})),
we obtain a relation;
\beq
G^1_N \gg \int dN' \;\; G^1_{N'} \;\; G^1_{(N-N')}.
\eeq
This relation reminds us of the  situation
in the two-dimensional quantum gravity (see, for example,
\cite{kawai}).
Two dimensional quantum gravity is known to describe a fractal space-time.
It consists of numerous tiny baby universes and
a single mother universe.
In our case, this analogy means that if we divide
any tree graph into two by cutting a bond,
we find only finite points in one of them and
the most of the points belong to the other.
In the case of $m$-point functions,
there is a single mother universe
on a blob in the path connecting these $m$ points.
Let's consider
the case of the two-point function as a simplest example.
Naively the figure \ref{fig:f2} implies that
the propagator behaves as that of the simpler random walk
if each blob is on an equal footing.
However one of the blobs in figure \ref{fig:f2}
becomes the mother universe consisting of infinitely many points
and it is entirely different from the other blobs.
Therefore we have to divide the path into two parts by cutting out the
mother universe.
Each part can be considered as a propagator of random walks whose
weights are dressed by blobs.
This is the reason why we have obtained
the propagator behaving as $1/p^4$ instead of $1/p^2$.
In other words, the mother universe (the blob with infinitely
many points) corresponds to a mass insertion and the other blobs
renormalize the correlation length $\xi$ of random walks from 
$a_0 N^{1/2}$ to $a_0 N^{1/4}$.
We can also apply a similar argument for the higher-point functions.
One of the blobs becomes the mother universe, which corresponds
to the single mass insertion in the $\phi^3$ scalar field theory. 
\par
\begin{center} \begin{large}
Acknowledgments
\end{large} \end{center}
We would like to thank Dr. S. Higuchi for correspondence.
This work is supported in part by the Grant-in-Aid for Scientific
Research from the Ministry of Education, Science and Culture of Japan
and by the National Science Foundation under Grant No. PHYS94-07194.

\section*{Appendix}
In this appendix, we derive the canonical-ensemble partition function
(\ref{eq:Zn}) from the Schwinger-Dyson equation (\ref{eq:sdb}).
Let us solve $b$ as a form of expansion in $\kappa$.
Each coefficient is calculated to be
\beqa
\frac{1}{2\pi i} \oint_{\kappa=0} d\kappa \frac{b}{\kappa^{N+1}}
&=& \frac{1}{2\pi i} \oint_{b=0} db (1-b) e^{-b}
\frac{b}{(b e^{-b})^{N+1}}\\
&=& \frac{N^{N-1}}{N!}.
\eeqa
Hence,
\beq
b=\sum_{N=1}^{\infty} \frac{N^{N-1}}{N!} \kappa^{N}.
\eeq
From the definitions of
(\ref{eq:cpf}), (\ref{eq:sdcf}) and (\ref{eq:defb}), $b$ is expanded as 
\beq
b=\hat{f}(0)\; \sum_{n=1}^{\infty} \frac{N}{V}\;
Z_N \; \kappa_0^N.
\eeq
Comparing these two expansions, we get the result of \eq{eq:Zn}.

\end{document}